RESEARCH ARTICLE

# The Landscape of Data Reuse in Interactive Information Retrieval: Motivations, Sources, and Evaluation of Reusability

Tianji Jiang[1] | Wenqi Li[2] | Jiqun Liu*[3]

[1] School of Education and Information Studies, University of California, Los Angeles, CA, United States

[2] Department of Information Management, Peking University, China

[3] School of Library and Information Studies, the University of Oklahoma, OK, United States

**Correspondence**
*: Corresponding Author: Jiqun Liu. Email: jiqunliu@ou.edu

**Abstract**

Sharing and reusing research data can effectively reduce redundant efforts in data collection and curation, especially for small labs and research teams conducting human-centered system research, and enhance the replicability of evaluation experiments. Building a sustainable data reuse process and culture relies on frameworks that encompass policies, standards, roles, and responsibilities, all of which must address the diverse needs of data providers, curators, and reusers.

To advance the knowledge and accumulate empirical understandings on data reuse, this study investigated the data reuse practices of experienced researchers from the area of Interactive Information Retrieval (IIR) studies, where data reuse has been strongly advocated but still remains a challenge. To enhance the knowledge on data reuse behavior and reusability assessment strategies within IIR community, we conducted 21 semi-structured in-depth interviews with IIR researchers from varying demographic backgrounds, institutions, and stages of careers on their motivations, experiences, and concerns over data reuse. We uncovered the reasons, strategies of reusability assessments, and challenges faced by data reusers within the field of IIR as they attempt to reuse researcher data in their studies. The empirical finding improves our understanding of researchers' motivations for reusing data, their approaches to discovering reusable research data, as well as their concerns and criteria for assessing data reusability, and also enriches the on-going discussions on evaluating user-generated data and research resources and promoting community-level data reuse culture and standards.

## 1 | INTRODUCTION

*Data reuse* refers to the use of data originally collected by others for a different research purpose from the original one (Zimmerman, 2008). This includes both re-using data to explore new questions or replicating previous studies. A recent multidisciplinary survey found that most researchers held a positive attitude towards data sharing and reuse, and over one-third of scientists regularly reuse others' data in their research (Tenopir et al., 2020). Proponents of data reuse believe it can boost research visibility, increase the efficiency and productivity of research investments, expose data to new tools, methods, and approaches, and foster greater global equality in scholarship (Borgman et al., 2015; Perrier et al., 2020). Despite these advantages, reusing others' data is challenging because it requires researchers to locate, access, and understand the data, often relying on metadata, documentation, contact with data creators, or their own expertise (Borgman, 2015; Pasquetto et al., 2019). The current digital ecosystem for data publications often falls short of meeting reusers' information needs (Engelhardt, 2022; Wilkinson et al., 2016), despite efforts to build infrastructures and progress in some fields (Klingner et al., 2023; Ugochukwu & Phillips, 2024).

Building infrastructures for research data reuse requires understandings of the diverse practices and needs of stakeholders, including data creators, curators, reusers, and research funders (Luesebrink et al., 2014). Many studies have investigated researchers' data reuse practices in different disciplines such as astronomy (Sands et al., 2012), archaeology (I. Faniel et al., 2013), and social sciences (R. G. Curty, 2016; I. M. Faniel et al., 2016). While these studies improved our understanding of the data reuse life cycle, variations remain in how data are shared, accessed, and reused across disciplines. Advancing data sharing and building infrastructures that support both intra- and cross-disciplinary exchanges relies on broader insights from researchers in diverse fields.

Interactive information retrieval (IIR) focuses on the study and evaluation of users' interactions with IR systems and their satisfaction with the retrieved information (Borlund, 2013), aiming to optimize retrieval processes based on empirical data (Zhai, 2020). As the field advances, the demand for large-scale and high-quality data continues to grow rapidly (Gäde et al., 2021), making data collection a time-consuming and resource-intensive task. Data reuse is considered a promising approach to overcoming these barriers by avoiding duplication of effort and improving result comparability. Efforts in IIR research include initiatives like the TREC Interactive (Over, 2001) and Session (Carterette et al., 2016) tracks, the INEX Interactive track (Nordlie & Pharo, 2012), and the Interactive Social Book Search track (Petras et al., 2019), along with efforts to standardize the collection, documentation, and sharing of IIR experiments and data (Gäde et al., 2021; Liu, 2022). However, data reuse practices within IIR remain unexplored, leading to a lack of empirical evidence for improving infrastructures for data sharing and reuse in this field. This study addresses this gap by investigating the experiences and perspectives of IIR researchers regarding data reuse. Through 21 semi-structured interviews with researchers across varying career stages and methodological orientations, we examine how researchers discover, assess, and reuse others' research data, as well as their motivations and concerns. Understanding the needs and challenges of data reusers is essential for improving the infrastructures that support data reuse, particularly within the IIR community, where methodological diversity and interdisciplinary collaboration create unique challenges. IIR research exhibits a large variety of research designs and methods (Kelly & Sugimoto, 2013), making it an ideal example to observe data reuse practices and challenges in a cross-disciplinary environment. Our findings will contribute to the development of infrastructures supporting data sharing and reuse across disciplines, benefiting a broader research community.

## 2 | RELATED WORK

Research data has become a public resource that is a "scientifically enlightened, morally worthy, politically progressive, and economically beneficial activity" (Mauthner, 2012). With the emphasis on open science and open data, researchers now have more propitious environment and better infrastructures to access and reuse datasets from previous research (R. Curty, 2015). Many disciplines have studied data reuse practices, including earthquake engineering (I. M. Faniel & Jacobsen, 2010), astronomy (Sands et al., 2012), medicine (Meystre et al., 2017), food science (Melero & Navarro-Molina, 2020), computer science (Koch et al., 2021), agricultural(Mwinami et al., 2022), and social science (Sun & Khoo, 2017). These work jointly revealed the data reuse landscape in the science communities.

### 2.1 | The motivations and concerns over data reuse

The benefits of data reuse are well recognized. A key motivation for data sharing is to avoid duplication of effort, thus accelerating science advances (Pasquetto et al., 2019; Rung & Brazma, 2013). For example, Yoon (2015) found that social work and public health researchers' primary motivations for data reuse is *cost-effectiveness* – they can focus on research questions without spending too much time on data collection. Data reuse can also expand research possibilities by providing researchers access to the data that are hard to collected by themselves (Kiesler & Schiffner, 2023). Further, data reuse supports new technologies like machine learning, which rely on large-scale data that is hard to obtain(Kölling & Utting, 2012), and addresses the "reproducibility crisis" by reanalyzing existing data to validate prior findings (Baker, 2016; Borgman et al., 2015).

Despite the potential benefits of data reuse and the growing consensus on sharing research data, many scientists still have concerns over data reuse. Bishop (Bishop, 2009) reported that qualitative researchers expressed concerns about potential ethical violations, given the direct interaction with human subjects in qualitative research. The boundaries between shareable research data and private information are often blurred, complicating the assessment of risks and responsibilities (Borgman, 2018). A DataONE survey (Tenopir et al., 2015) revealed that while researchers value data sharing and reuse, there were also increased concerns about the risks of misuse and misinterpretation. Data are embedded in a local context, and removing data from their original context inevitably leads to information loss (Pasquetto et al., 2019). Communicating the contextual information to the reusers was found to be a big challenge across disciplines (Borgman & Groth, 2024; I. Faniel et al., 2013; I. M. Faniel & Zimmerman, 2011), and the loss of such information limits reusers' understanding of the data. Furthermore, reused data are sometimes perceived to be less valuable than originally collected data, since they require less effort to produce or collect (R. Curty, 2015). Researchers may worry that reusing others' data will devalue their work and reduce publication opportunities (Yoon & Kim, 2017).

## 2.2 | Factors influencing data reuse practices

Factors influencing data reuse practices have been identified in previous studies across disciplines. Tenopir et al. (2015) found in an interdisciplinary survey that the availability of contextual information, as well as the quality of and trust in data, played important roles in researchers' decisions regarding data reuse. R. G. Curty (2016) identified six categories of factors influencing social scientists' data reuse experiences: perceived benefits, perceived risks, perceived efforts, reusability assessment, enabling factors, and social factors. Data quality metrics, including completeness, accessibility, ease of operation, and credibility, were found important in the assessment of data reusability (I. M. Faniel et al., 2016). Regarding the information loss, Borgman and Groth (2024) identified six dimensions of distance between reusers and the data's original context: domain, methods, collaboration, curation, purposes, and time and temporality.

Pasquetto et al. (2017) investigated astronomers' data reuse practices, including the needs of data reuse, motivations for data reuse, and the factors that shape opportunities for reusing data. Wang et al. (2021) reviewed 42 recent studies on data reuse and identified three key stages: initiation, exploration and collection, and repurposing. They also explained how these stages interact and exhibit iterative characteristics. Surveys by researchers (K. Gregory et al., 2020; Joo et al., 2017) further revealed factors shaping decisions and behaviors in data reuse.

## 2.3 | Data reuse practice in IIR

IIR researchers, like those in many other fields, rely on empirical data for their studies. As research in IIR advances, the need for large-scale, high-quality data continues to grow(Gäde et al., 2021), making data collection (especially human feedback labels) a time-consuming and resource-intensive task. The benefits and importance of reusing research resources has long been recognized in the IIR community. Several initiatives have addressed challenges in collecting, managing, preserving, and sharing IIR research materials, including the Text Retrieval Conference (TREC) Interactive (Over, 2001) and Session (Carterette et al., 2016) Tracks, the INEX Interactive Track (iTrack) (Pharo et al., 2010) , Cultural Heritage in CLEF (CHiC) Interactive Task Petras et al., 2013, the Repository of Assigned Search Tasks (RepAST) (Freund & Wildemuth, 2014), and the INEX 2014 Social Book Search Track (Bellot et al., 2014). These efforts provided valuable insights and firsthand experiences on the challenges and opportunities associated with sharing and reusing IIR research materials.

Researchers have also studied resource reuse in IIR and attempted and madetheoretical contributions and provide suggestions future practices. Issues related to re-use in IIR have been discussed at various workshops, such as the Supporting Complex Search Tasks (SCST) workshops in 2017 (Belkin et al., 2017) and the Workshop on Barriers to Interactive IR Resources Re-use (BIIRRR) in 2019 (Bogers et al., 2019). These workshops discussed reuse experiences and the need for better documentation to facilitate future reuse. Additional research has focused on evaluating and improving the reusability of IIR resources. Liu and Shah (2019) developed a framework that presents a series of user study facets (e.g. participant population, study protocol, experimental system, methods of analysis) to guide IR research documentation and evaluation on the reusability of IIR research resources. Gäde et al. (2021) discussed the challenges in documenting and archiving the contextual information of research resources generated in IR studies, and proposed eight principles for improving the reuse of IR research resources. Liu (2022) proposed a Cranfield-inspired reusability assessment framework to assess the reusability of IIR research resources and examine experimental reproducibility.

In the IIR community, task-related resources (both work task and search task) have received more attention than other types of resources, such as research data(Liu, 2022). Previous studies on data reuse often provide a field-specific perspective, limiting their generalizability to the unique and broader contexts of IIR. IIR researchers may have different attitudes and practices from other fields. Therefore, it is important to see how IIR researchers reuse data, their concerns, and how these differ from that of other fields.

## 3 | RESEARCH QUESTIONS

This interview-based study is developed as part of a larger research agenda aimed at understanding data sharing, curation, and reuse practices within the IIR community. Our focus here is on the current state of data reuse behaviors within the community, as well as the factors influencing data reusability from the perspective of data reusers. The term 'reuse' refers to the use of existing data that was gathered for a previous research study for purposes that are different from the original intention of the data producer (Zimmerman, 2008). 'Data reuser' refers to the person who attempts to reuse such data in their research activities. 'Data reusability' refers to the perceived value of existing data by reusers or potential reusers for secondary analysis to address new questions. The study's objective is to gain insights into IIR researchers' data reuse behaviors, including their motivations, attitudes and reuse experiences, and challenges they face when attempting to reuse data in their studies. The following four questions guided our study:

1. What are the main intents/motivations for researchers to reuse data?
2. How do IIR researchers discover and obtain reusable research data in their research practices?
3. How do IIR researchers assess the reusability of research data shared by others?
4. What are the main concerns that demotivate IIR researchers from reusing others' data?

## 4 | METHODS

Survey and interview represent the two most commonly used methods in examining data sharing and reuse. Survey research enables researchers to identify common practices, behaviors, and perceptions related to data sharing and reuse within a target population. It also allows for the exploration of factors correlated with data sharing and reuse, and provides quantifiable insights into how these factors may influence these behaviors. (I. M. Faniel et al., 2016; Kim & Nah, 2018; Melero & Navarro-Molina, 2020). Interviews, from an alternative approach, allows researchers to gain in-depth understandings of the thoughts, feelings, and perceptions of subjects from their point of view with their own words (Pickard, 2013). Going beyond general trends or patterns, an interview study involves researchers actively seeking out and exploring distinct individual experiences, which contribute to a richer and deeper understanding of data sharing and reuse. (Borgman, 2015; Borgman et al., 2019; K. Gregory et al., 2020). Moreover, interviews allow researchers to learn about participants' firsthand experiences and engage in real-time interactions with them. This enables researchers to clarify any confusion and delve deeper into interesting points raised by the participants. As a result, interviews provide more accurate and in-depth insights directly from study participants, ensuring a reality-based understanding of their perspectives and experiences. Interview also allows researchers to identify subtle insights and complexities that are often overlooked or difficult to capture through survey-based methods, making it particularly valuable for exploring nuanced behaviors and motivations. To have a accurate, broad, and in-depth understanding of the data reuse practices of IIR researchers, we conducted semi-structured interviews (N = 21) in this study to address our research questions. We followed the interview process recommended by Pickard (2013). Figure- 1 illustrates the research process.

### 4.1 | Interviews

All interviews followed a predeveloped protocol. Non-directive, open-ended questions were asked about the participants' profile information and data reuse experiences, focusing on the central themes we would like to address in this study, including the reasons that motivate them to reuse others' data, their decision-making process when weighing the tradeoff between reusing others' data and collecting their own, the obstacles encountered during such experiences, and their resolutions. They were encouraged to share their experiences with real cases to help us have a better understanding of their points. We then asked participants to step outside their specific cases of data reuse practices and responded to a few additional open-ended questions about their perspectives on data reuse as a research method in IIR studies, including the benefits and advantages, potential risks and concerns, and scenarios in which data reuse might be appropriately for. Based on the participants' responses, we followed up with additional questions to seek clarification or gather more information. Before conducting the interviews, we held 2 pre-interviews with our colleagues to validate the interview process and refine the questions to better address our objectives.

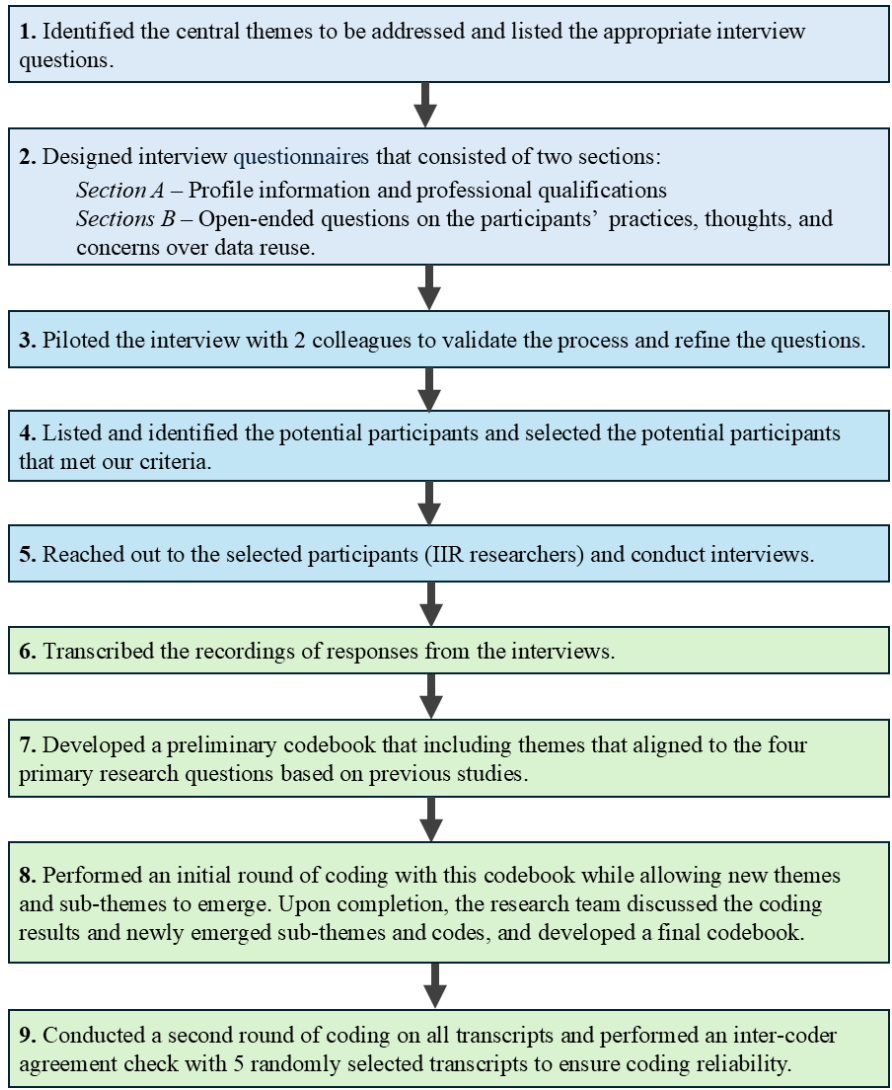

**FIGURE 1** Flow of data collection and analysis

Following the interview protocol, we conducted 21 semi-structured interviews via video conference tools including Zoom and Voov, according to the preference of participants. To avoid confusion from the varying definitions of "reusability" in prior studies, we adopted the definition of "reusability" from several influential works (I. M. Faniel & Jacobsen, 2010; Meystre et al., 2017; Pasquetto et al., 2017, 2019; Wallis et al., 2013): *the degree to which research data is able or fit to be used for purposes different from the one for which it was originally collected in later studies.* The definition was presented to participants at the beginning of each interview. The duration of the interviews ranged from 30 to 60 minutes, with an average of around 40 minutes.

## 4.2 | Participants

The study participants were recruited through purposeful sampling. We deliberately selected a cohort of researchers with a minimum of 3-year experience in conducting IIR research and experiences in attempting to reuse other's data in prior research endeavors. Recognizing the existence of two distinct communities among IIR researchers with respect to their research methodologies and topics – the system-oriented researchers concentrating on data-driven experiments, offline measures and algorithms, and the user-oriented researchers studying and evaluating IIR systems from user perspectives – we endeavored to achieve a balanced representation of participants from both communities. This approach helped minimize biases related to research assumptions, methodologies, and interpretations. After receiving IRB approval from the first author's institution, we initiated recruitment through our professional network and expanded the pool using snowball sampling, where our study participants referred additional appropriate candidates. Recruitment continued until we reached theoretical saturation.

In total, we recruited 21 participants, comprising 12 user-oriented and 9 system-oriented IR researchers. The group included 8 professors, 1 post-doctoral researcher, and 12 senior Ph.D. students, all of whom had rich experiences in the IIR research. Seventeen participants had reused data from colleagues, while the remaining four, who had considered data reuse, shared their reasons for not proceeding. Despite the relatively small sample size, the participants still represented a wide range of backgrounds and experiences, specific research focuses within IR, stage of researcher career, and geographical locations. We present additional participant demographic information in Table 1.

**TABLE 1** Demographic Information of the Participants

| Participant | Position | Place of Affiliation | IIR Research Interests |
| --- | --- | --- | --- |
| Participant 1 | PhD Student | Japan | recommender systems, metrics for IR system evaluation |
| Participant 2 | Professor | United States | users and interactive retrieval |
| Participant 3 | PhD Student | China | conversational IR systems evaluation |
| Participant 4 | Post-doc | United Kingdom | task-oriented conversational systems |
| Participant 5 | PhD Student | China | learning-related searching behaviors |
| Participant 6 | PhD Student | United States | human search behaviors, human-computer interaction |
| Participant 7 | Professor | Japan | search behaviors, IR model |
| Participant 8 | PhD Student | United States | health information retrieval behaviors |
| Participant 9 | Professor | United States | human search behaviors |
| Participant 10 | Professor | China | IR user behavior, metrics for IR system evaluation |
| Participant 11 | PhD Student | United States | human-computer interaction |
| Participant 12 | PhD Student | United States | human search behaviors |
| Participant 13 | PhD Student | United States | IR user behavior, biases in IR system |
| Participant 14 | PhD Student | China | human search behaviors, human-computer interaction |
| Participant 15 | PhD Student | China | learning-related searching behaviors, search as learning |
| Participant 16 | PhD Student | China | human interactions with AI searching tools |
| Participant 17 | Professor | China | user behavior analysis, retrieval models |
| Participant 18 | Professor | United States | IR systems evaluation |
| Participant 19 | Professor | United Kingdom | recommender systems, IR system |
| Participant 20 | PhD Student | Australia | human search behaviors |
| Participant 21 | Professor | Australia | conversational search, human-computer interaction |

## 4.3 | Analysis

The interviews were recorded and fully transcribed with the permission from participants. They were then analyzed using a qualitative data analysis tool, MAXQDA 2022. We conducted two rounds of transcript coding. Based on findings from previous studies R. G. Curty, 2016; I. M. Faniel et al., 2016; Liu, 2022, the first author developed a preliminary codebook including themes that aligned to the four primary research questions. These themes were related to the purposes and motivations for data reuse (RQ1), the discovery and access of others' data (RQ2), the assessment of data reusability (RQ3), and the concerns and challenges over data reuse (RQ4). The first and second authors then performed an initial round of coding with this codebook while allowing new themes and sub-themes to emerge. The two authors first selected five transcripts with rich information and coded them independently. Upon completion, the entire research team convened to discuss and compare the coding results and newly emerged sub-themes and codes. This collaborative process led to the refinement and consolidation of themes, sub-themes, and codes, resulting in a final codebook (see Supplementary Materials). Involving more people in this process helped reduce bias from the first author who created the preliminary codebook, as well as bring new insights to the codebook. Using the finalized codebook, the first author conducted a second round of coding on all transcripts to ensure consistency. The transcripts were iteratively coded with themes, sub-themes, and codes such as purposes and motivations (e.g., answer new research questions based on the data), discovery and access of other's data (e.g., learn about reusable data through personal connections), and concerns and challenges of reuse (e.g., non-standardized or inaccurate documentation). To ensure the reliability of the coding results, the second author independently coded a randomly selected sample of three transcripts. Inter-coder agreement was assessed using Cohen's kappa (Brennan & Prediger, 1981), a statistical measure widely used to assess consistency between coders. The kappa coefficient exceeded 0.6, which is generally interpreted as substantial agreement. This level of agreement confirms that the coding framework was applied consistently and suggests that the first author's coding results of the transcripts are reliable.

## 5 | RESULTS

Overall, our findings revealed different data reuse practices between system-oriented and user-oriented IIR researchers. All the system-oriented participants reported frequent data reuse practices in their studies, typically using data created by individuals or institutions with whom they had no direct connection. In contrast, user-oriented participants reused data less frequently, often within their own research teams or collaboration networks. Yet, the two participant groups jointly represent a broad, diverse set of data reuse and reusability evaluation strategies.

We present the findings in four parts parallel to our research questions. The first part describes the participants' purposes and motivations for reusing data (RQ1). The second explores how participants discover and access reusable data (RQ2). The third discusses their strategies for assessing data reusability (RQ3). Finally, the fourth investigates challenges and concerns expressed about data reuse (RQ4).

### 5.1 | Purposes and motivations of IIR researchers' data reuse

There was no universal definition of data reuse among study participants. While using data collected by others for different purpose is widely recognized as data reuse, participants sometimes included cases that were not strictly qualify as "reuse". These include continuing the analysis of data collected by lab colleagues in ongoing projects, using data originally gathered for non-research purposes, or repurposing their own data for entirely different goals. Overall, participants reused data for two main purposes: exploration and ground-truthing. Exploration supports researchers in discovering new insights or patterns, while ground-truthing involves testing hypotheses, generalizing findings, or replicating results. It's worth nothing that the user-oriented study participants primarily reused data for exploration, rarely for ground-truthing.

In practice, participants treated reused data similarly to their own data but with some differences in reporting. The main difference lies in how they reported their research design and data collection in the study papers. When discussing their own data, participants typically described the data collection process and provided examples to help readers understand it. For reused data, however, participants focused on defending their choice, justifying their understanding, and demonstrating effective use. They usually did not elaborate on how the data was originally created. Instead, they would mention the source of the data and provide citations to the original study for which the data was created.

Participants highlighted several benefits of data reuse: saving time and resources, accessing otherwise unavailable data, increasing study reliability, enabling comparability with prior research, and fostering new collaborations. The motivations are based on the IIR researchers' research orientation and methodologies. For example, participants focusing on IR system evaluation often mention that reusing external data can reduce biases and avoid manipulation in data by applying their research findings to different datasets. They also need colleagues' data to test if their research findings are applicable to broader settings. A participant focusing on developing IR system mentioned a different motivation - *"reusing data is the only way for a junior researcher to access the data needed for research ... large-scale data is necessary for developing IR system, but such data are difficult and expensive to collect."*

## 5.2 | Discovering and accessing reusable research data

### 5.2.1 | Discovering data

Participants typically discovered reusable data through their own research experiences or their personal connections, such as academic advisors, colleagues, or collaborators. When conducting research, they might recall data encountered earlier and evaluate its suitability for current research projects. If they couldn't recall any data that might be reusable for the ongoing study, they would collect their own data. They rarely conduct active search for reusable data during this process. This finding aligns with previous studies in other fields(I. M. Faniel & Jacobsen, 2010; Kim & Yoon, 2017).

Academic literature, especially journal articles and conference papers, was reported as the most common source for discovering datasets. For system-oriented IIR researchers, conferences and workshops are also frequently mentioned sources. Participants had various approaches to discover and learn about data through academic literature. If the literature was based on originally collected data, they usually learn about the data through methods and data description sections in publications or check the supplementary materials to see if the data is relevant. However, most participants stated they wouldn't examine the original data, unless they had strong suspicions about the data descriptions in the literature. On the other hand, if the literature was based on reusing others' data, more common among system-oriented participant, participants would learn about the data through words introducing and defending the selection of the data source. If they found the data interesting, they would trace citations to original datasets. For example, a participant noted:

> *Reading academic papers is definitely the most common way (to find reusable data)... I mainly (discover data by) read(ing) research papers, and from there, I gain an understanding of what kind of data others are using, what the data was about, and what could be done (with the data) ...The paper can also direct me to the data sources.*

Personal connections were reported as the most effective way to identify and access reusable data, such as direct communication with academic advisors or colleagues and attending workshops or lectures, or through virtual academic communications on social media platform like Twitter. Almost all of the participants mentioned experiences of discovering novel data through personal connections, either by chance or through intentional efforts. And most of the participants who reused data in their studies reported that the data they reused was discovered through this approach in their data reuse practices. Information about data received through personal connections is believed to be more accurate and trustworthy than that obtained through other approaches. Furthermore, it allows individuals to directly contact those who have collected the data or possess in-depth knowledge of it, making the data easier to understand and access. As a participant described:

> *When I was discussing my research with my advisor, she mentioned that a senior student in our lab had collected a dataset that could be useful and even showed me a sample of the data ... I reached out to the student and she agreed to share the dataset with me ... Whenever I had questions about the dataset, I would directly contact the student, and she would explain it to me. Without her assistance, I wouldn't have been able to understand the dataset.*

Only a few system-oriented participants studying information retrieval systems used public repositories or search engines to discover reusable data. Repositories like TREC and MS MARCO were commonly mentioned by system-oriented participants, but cross-disciplinary platforms were rarely used, indicating a gap in awareness or adoption despite the growing interest in developing and deploying such infrastructures in the recent years (Chapman et al., 2020). This discrepancy highlights the need to bridge the gap between the infrastructure developers and end-users, and emphasizes the importance of educational initiatives to inform researchers about the available tools and their benefits.

### 5.2.2 | Accessing data

Accessing data relied heavily on personal connections. In particular, all user-oriented participants' successful data reuse practices involved interactions with individuals they knew. There are several reasons for this. First, data used in IIR studies often involve human subjects, and storing these data on publicly accessible platforms may raise privacy concerns. Second, to reduce the risk of data misuse, data providers often prefer direct contact with researchers to clarify reuse purposes and grant access accordingly. Third, The data providers would like to seek new collaboration opportunities through this reuse of their data. Additionally, as observed in other fields (Lopez-Veyna et al., 2012), data are considered valuable and exclusive research resources, and access is typically restricted to a small, trusted circle of collaborators. A participant described his experience reusing data in collaboration with a company research team: *"I have to be one of 'their people' to access the data ... Without the collaboration with the company, I wouldn't even have been aware of the existence of this data."*

Lab websites, such as the Information Retrieval Lab at Tsinghua University (THUIR), and personal repositories like GitHub and Google Drive were more commonly used to share large-scale machine-generated data that has minimal privacy or ethical concerns. Data providers often share links to their data in published literature, sometimes with accompanying documentation. Compared to accessing data via personal connections, this approach was considered more convenient for data reusers while still allowing data providers to maintain full control over their data. Data providers can keep their data unpublished or edit it at any time and monitor its usage.

## 5.3 | Assessing the reusability of other's research data

Our findings revealed four key issues in assessing data reusability: understandability, trustworthiness, previous usages, and data collection methods. Participants also discussed the contextual information and strategies they employed to make the assessment. Table 2 presents a summary of the information required and the resources employed by IIR researchers in their evaluation of data reusability.

### 5.3.1 | Understandability

Understandability of data refers to how easily a researcher can make sense of the context, structure, content, and other essential information of a dataset. When participants attempted to reuse a dataset, they needed to figure out *'what information the dataset is conveying to researchers'* and *'how the information is transcribed into the text or numbers (or other symbols) they see in the dataset'* (participant 4). This process often required more than reading the data and associated documents. Participants had to gather information from every available source to thoroughly comprehend the logistics and meanings of all the variables, and then determine which portion of the data could be reused and how. Missing or undocumented information sometimes made reuse impossible. As one participant described:

> *Understanding all the variables in a dataset is super challenging, and sometimes even impossible without direct assistance from the data collector ... Once I was reading a variable about users' satisfaction with a search interface. Even though I knew that the values of the variable were provided by research participants using a seven-point Likert scale, I still couldn't understand what the values exactly meant because I didn't know how 'satisfactory' and 'unsatisfactory' were defined in the survey, and what the distance was between each level of satisfaction.*

### 5.3.2 | Trustworthiness

Trustworthiness depends on the reliability and validity of the dataset. Participants assessed these factors by examining how data were produced and used previously.

To assess dataset reliability, researchers typically examine factors such as data collection methods, instruments used, and bias reduction techniques. However, for IIR researchers, this approach is often impractical for two reasons. First, essential information is frequently inaccessible. Unlike well-documented fields such as physics or astronomy, IIR data often involve significant researcher intervention, requiring detailed disclosures that are rarely provided. Contextual information provided in research papers often lacks sufficient depth for verification, with many details undocumented. Second, even when documentation is available, its accuracy and completeness cannot always be assured, as there is no standardized framework for documenting IIR data collection. These challenges are particularly prominent in user-oriented IIR studies, where human intervention and diverse research designs exacerbate difficulties in verifying data production.

**TABLE 2** A summary of IIR researchers' reusability assessment strategies

| Assessment | Metadata Needed | Sources |
|---|---|---|
| Can data be understood? | Information about data collection (experimental setting, task, data collection method); Definitions of variables; Description of data structure; Information about data processing; | Complementary documentation; Metadata provided by data repository; Conversations with colleagues; Papers introducing the data; |
| Are datasets trustworthy? | Academic credentials of the data collectors/producers; Information about prior (re)use; Type of data; | Conversations with colleagues; Personal networks; |
| What can be done with the data? | Purpose of collecting the data; Information about prior (re)use; | Papers introducing the data; Papers (re)using the data for research; |

To answer "is the data reliable" question, participants commonly assessed the academic reputation of individuals or institutions associated with the dataset. A dataset is considered reliable if it was produced or curated by *"a person or institution that has a good reputation or associated with someone I know that have a good academic records"* (Participant 3). Conversely, data from unfamiliar institutions raised concerns, limiting reuse to trusted networks. This pattern reflects a "Matthew Effect" in data reuse, where datasets associated with well-known figures tend to receive greater recognition for their reliability.

Validity is another major concern when participants assess the trustworthiness of a dataset, and their strategies differ by research orientation. For user-oriented participants, assessing validity typically involves examining data collection design, including sample size, participant diversity, selection criteria, and experimental environment (Liu, 2022). They evaluate whether these elements allow data producers to collect the intended information. System-oriented researchers, by contrast, assess validity not only by examining how the data was collected but also by reviewing its prior usages. A dataset is often considered valid if referenced in numerous or influential studies. This is because these researchers prioritize comparability with prior research. Thus, if the data has been previously used, especially in studies they aim to compare with, they are less concerned about its validity.

### 5.3.3 | Previous usage

Researchers also consider the previous usage of data when assessing reusability. Generally, they believed that datasets serving a wide range of purposes and adapting to various circumstances are more reusable. If a dataset has been reused in previous studies, those prior usages provide evidence to justify its reuse in new research. When assessing a dataset's reusability for a specific research project, participants did not merely count how many times the dataset has been used before. Instead, they carefully examine each instance of reuse, evaluating the purpose and method to see if it matches their current needs. They prefer datasets that were collected for research purposes similar to their own. Typically, participants begin by reviewing the original research papers for which the dataset was created. They then used it as a starting point to trace subsequent research papers that reused the data.

### 5.3.4 | Data collection methods

Information about data collection was important for most participants when evaluating data reusability. A participant emphasized, *"I have to understand all the details of the actual data collection process in order to make proper use of the data."* The participants expressed a strong desire to comprehend *"exactly how the data was collected"* and *"all the details of the actual experiment that collected the data"*. They sought to understand the experiment's complexities as if they had been present during its execution. Understanding these details is critical for assessing whether the dataset fits their specific research requirements. Without comprehensive information about how and for what the dataset was gathered, participants faced challenges in evaluating whether the data collection design matched their research questions and objectives. As a result, when participants lacked this information, they were often reluctant to reuse the data due to concerns about potential discrepancies or limitations that could affect the validity of their research.

## 5.4 | Concerns over data reuse

Although most participants held a positive attitude towards data reuse, they also mentioned issues that demotivate them in their practices. These concerns corresponds to different steps involved in the data reuse process (Table 3 ).

### 5.4.1 | Understandability concerns

Challenges in understanding other's data were a major concern for participants regrading data reuse. These challenges often stemmed from absence of key contextual information, including data collection design, participant selection criteria, quality control methods, variable definitions, equipment details, measurement techniques, and the verification of data integrity.

Absence of contextual information arose from either objective or subjective reasons. Objectively, this information might not have been included during data-sharing, either due to producers failing to capture it during the data collection or data sharers omitting details in related papers or documentation. Subjectively, even when such information was included, researchers might struggle to understand the process or trust its accuracy due to vague, non-standard, or unfamiliar presentation.

The difficulties in understanding others' data were also rooted in participants' research orientation. User-oriented participants often faced greater uncertainties, even after going through the raw data and related documentation, due to absent details like participant selection, task sequences, or real-time feedback participants. Their uncertainty also stemmed from doubts about whether the data were collected rigorously and in accordance with the procedures described in the documentation. In contrast, system-oriented participants reported fewer subjective challenges that hindered their understanding of others' data. Their datasets were simpler, with less dependence on detailed contextual information. Instead, their difficulties were primarily objective, such as interpreting variable meanings. The documentation usually provided little information about the variables' definitions and the range of values for each variable.

### 5.4.2 | Usefulness concerns

Perceived usefulness of data was another frequently mentioned concern discouraging data reuse. This concern had two aspects: (1) others' data might not align with their research goals; (2) the data might already be fully analyzed by the original collectors, leaving limited opportunities for innovative studies.

Usefulness concern was especially common among user-oriented IIR participants. For them, data was typically collected by observing users interacting with systems to complete tasks in specific settings, using methods like experiments, interviews, or surveys. The design and methods of data collection were closely tied to specific research questions, making it challenging for researchers with different questions to re-purpose the data. Additionally, collecting such data required significant resources and time. It was usually only useful for narrow research fields that the data collectors themselves were working on. Therefore researchers who originally collected the data tended to maximize its use before sharing it with others, leaving little room for innovative studies reusing the data.

In addition to common factors that concern researchers in other fields, such as reliability (Pasquetto et al., 2019), validity (I. M. Faniel & Jacobsen, 2010), and legal constraints (Melero & Navarro-Molina, 2020), our participants highlighted unique requirements: timeliness and privacy concerns. IIR focuses on user interaction with retrieval systems, emphasizing user behavior, experience, and cognitive processes (Liu, 2022). With rapidly evolving technologies, user behavior also changes quickly. As a result, our participants were particularly concerned with the timeliness of their data. As a participant stated, *"To develop retrieval tools that meet current needs, researchers must conduct studies based on data that reflects current user behavior, which places high demands on the timeliness of the data."* Furthermore, since IIR data often contain sensitive personal information, such as identification details and search history, participants were highly concerned about privacy whether in the collection, archiving, sharing of their own data, or in the reuse of others' data.

### 5.4.3 | Social or community related concerns

Another concern raised by participants was the acceptance of data reuse within their research community. While participants generally held positive attitudes toward data reuse and stated they would not undervalue studies based on reused data, they worried that others might not recognize such work. These concerns centered on two perceived shortcomings: reliability and innovation.

**TABLE 3** Concerns over data reuse for IIR researchers: Key points

| Understanding other's data (Understandability concerns) | Evaluating usefulness of other's data (Usefulness concerns) | Attitudes of the research community (Social/Community-related Concerns) |
|---|---|---|
| Critical contextual information about a dataset is sometimes missing when it is shared with other researchers. | The datasets shared by others usually do not fit the data reusers' needs. | The study participants worry that research based on reused data will be devalued or receive many doubts from their colleagues. |
| Data reusers have difficulty interpreting or gaining trust in a dataset and the contextual information provided by other researchers. | Researchers who originally collected the data tend to maximize its use before sharing it, leaving little room for innovative studies using the reused data. | If a study is based on reused data, its reliability is largely beyond the reuser's control and depends on the reputation of the original data collectors and any previous studies that used the data. |
| | The data collected by others may lose useful information due to the passage of time or privacy issues. | If a study is based on reused data, it may be devalued by peer reviewers as lacking innovation in research design. |

The reliability of a study's findings heavily depends on its data collection process. For studies using reused data, reliability partly rests on the reputation of the original data collectors and previous studies using the same data. This may not be a major issue when data providers are well-known or reputable or if the data has been widely cited. However, if these conditions are absent, reviewers might question the data collection process, leaving reusers unable to defend it. As a participant said *"reusing other's data means the reliability of your research is vouched for by someone else...can be risky."* This partly explains why participants limited data reuse to within their labs or trusted research communities.

Concerns about the lack of innovation were another issue that discouraged data reuse, particularly among user-oriented participants. Participants speculated that this might be due to the nature of their studies: user-oriented research often involves a big variety of methods and design for collecting data from specific information users. The studies could have many innovations in data collection design but typically less variety in data analysis methods, leaving fewer opportunities for innovation in reused data. As a participant said *"The design of data collection usually contributes a lot to the innovations of a study, and thus a study might seem less innovative if the data collection design is not present."*

## 6 | DISCUSSION

To inform the future development of infrastructures supporting data sharing and reuse in IIR community, we focus on illustrating how the researchers reuse other's data for their studies, how they assess the reusability of other's data, as well as their thoughts and concerns over it. We believe that our findings on data reuse practices, strategies for assessing data reusability, and concerns about data reuse offer valuable insights for advancing studies and practices aimed at facilitating data sharing in IIR. These findings may also inspire research on data sharing and reuse in other interdisciplinary fields and provide implications for the development of cross-disciplinary data-sharing infrastructures.

Compared to fields investigated in the previous studies on data reuse, such as social science (I. M. Faniel et al., 2016) (R. G. Curty, 2016), humanities (Li et al., 2024), earthquake engineering (I. M. Faniel & Jacobsen, 2010), astronomy (Sands et al., 2012), and clinical research (Meystre et al., 2017), IIR is more multidisciplinary. As a recently emerged interdisciplinary field, IIR attracts researchers from a wide range of disciplines, including computer science, library and information studies, psychology, and others. This diversity in backgrounds suggests that researchers in IIR are trained under different research paradigms, which in turn leads to different emphases and orientations in their exploration of the communication system between users and the IR system. For example, participants focused on system-oriented IIR research predominantly come from computer science backgrounds. It also leads the IIR researchers with different disciplinary background to hold differing views on many data-related issues, such as what constitutes usable data, how data should be collected and analyzed for research, how data should be curated and shared after research, and how to verify the reliability and validity of other's data.

Furthermore, according to the study participants, most of their research collaborations and data reuse practices occurred with researchers who shared the same research orientation. It is still uncommon for them to cross the boundaries and collaborate with someone who conducted IIR research in a different way. This makes data reuse practices among IIR researchers exhibit some unique characteristics compared to fields dominated by single-discipline researchers or fields where interdisciplinary collaboration is common. Furthermore, the multidisciplinarity makes the IIR field an interesting case for studying data sharing and reuse behavior in an environment closer to the reality, which may provide some insights for future studies on multidisciplinary or even interdisciplinary environment.

Data reuse is a common practice among IIR researchers with the same research orientations but remains rare among researchers with different. In addition, the motivations, practices, and concerns over data reuse are also different between user-oriented and system-oriented IIR researchers. For user-oriented researchers, the common motivation for reusing data in their studies is further exploration. In contrast, system-oriented researchers typically reuse data to verify and generalize their findings or to make their work comparable with other studies based on the same data. Regarding concerns about data reuse, user-oriented IIR researchers express more worry about the loss of key contextual information compared to their system-oriented counterparts. They also raise more concerns about the acceptance of data reuse within their research community and ethical issues. Practices for assessing the reusability of colleagues' data also differ between IIR researchers with different focuses. System-oriented researchers place more emphasis on the trustworthiness of the data, while user-oriented researchers focus on its understandability.

Considerable information loss occurs when data are removed from context and shared between researchers Borgman and Groth, 2024, even though efforts have been made to archive and share some of the context of data. Data reusers must address this information loss before reusing the data, and a certain portion of this loss is unrecoverable, requiring the reusers to make assumptions to fill in the gaps. The assumptions are not based on mere speculation; rather, they are informed by the data reuser's familiarity with the data creators and the original purpose for which the data was created. The greater the familiarity, the fewer and more confident the assumptions data reusers will make. If data reusers have less familiarity, they will need to make numerous and potentially dubious assumptions, which will finally discourage them from reusing the data. This study supported the argument that participants felt more confident interpreting, evaluating, and reusing data collected by researchers who shared similar training and research paradigms (e.g. their labmates, people sharing disciplinary backgrounds with them). Furthermore, if researchers in a field are not very familiar with each other and their work, the field may develop a generally suspicious attitude towards studies based on data reuse, which, in turn, may further discourage researchers from attempting to reuse data in the field. As we found with the study participants conducting user-oriented IIR studies involving user experiments, their less optimistic view of data reuse was due to the lack of standardized methodologies for conducting such experiments. Participants noted that researchers often designed and conducted experiments based on their own standards, making it difficult to assume the context of data which were not shared.

Increasing familiarity with others' work will enable researchers who attempt to reuse data to make fewer and more confident assumptions about the data's context. From our study, the familiarity can be gained from two levels: the individual level, and the research-community level. At an individual level, data reusers become familiar with others' work through personal experiences, such as reading publications, attending workshops or lectures, recognizing others' reputations, or engaging in direct communications. The approach effective but usually restricted within a relatively small research network. At the research-community level, familiarity depends on whether and how much the research community has a consensus on the design and methods regarding data collection, processing, documentation, and curation. For example, system-oriented IIR researchers mentioned that they typically do not question the trustworthiness and context of data from TREC, because they are confident that the data were collected, processed, and documented according to standards commonly accepted by their research community. As a result, there is no suspicion over the data itself raised by others in their field about the data when it is reused for study. Our study found that the IIR community primarily relies on individual-level approaches to become familiar with others' work. Most data reuse practices among our participants occur with individuals closely related to them, such as labmates or academic advisors. Research-community level approaches are adopted only by researchers in a few subfields, such as IR system evaluation. In most IIR subfields, and for the broader IIR community, researchers rarely become familiar with others' work through research-community level approaches and often lack the necessary infrastructure to support such practices. As a result, data reuse within IIR exists but is limited in scope. Data reuse across subfields is still quite rare.

Based on our findings, a major objective for motivating and facilitating data reuse in IIR should be to bridge the gap between data creators and data reusers. An effective approach would be to develop infrastructures that enable the IIR research community to become familiar with others' work through research-community level approaches. This could involve creating commonly accepted frameworks for reporting research design and data collection in publications relevant to the original study or establishing standards for collecting and processing data. Some researchers have already made contributions in this direction (Liu, 2022), and we anticipate further exploration and practices by the IIR community in the future.

Furthermore, The majority of interviewed IIR researchers do not habitually engage in active searches for research data. Instead, they rely on passive methods for gathering pertinent information, such as reading academic literature, seeking recommendations from colleagues, attending field-related lectures, workshops, and conferences. Nearly all participants indicated their lack of familiarity with data repositories, data search engines, and open data platforms. However, when discussing the primary challenges encountered while attempting to reuse data for research, most participants identified the significant obstacle of not being aware of available sources of reusable data. This discovery serves as an inspiration for future initiatives aimed at promoting data sharing and reuse in the IIR field. On one hand, we should enhance training in data search and data repository utilization in research education and activities within this field, enabling researchers to shift from their traditional data search habits and develop the skills needed to locate accessible data sources. On the other hand, as we develop future data retrieval tools, we should also develop data recommendation systems tailored to users' passive information acquisition tendencies. Such systems would assist scholars in discovering data that is both interesting and valuable for reuse.

## 6.1 | Limitations

This study has potential limitations. First, the findings are based on the participants' self-reported thoughts and behaviors regarding data reuse. However, we recognize the possibility that these reports may differ from the participants' actual practices in their research. In future studies, we plan to conduct on-site observations of IIR researchers' data discovery and reuse behaviors to gain a better understanding of their attitudes and practices towards data reuse. Second, as the study adopts a semi-structured approach, the conversations with participants may be influenced by the interviewer's interests, which could introduce biases and affect our findings. Third, the study participants demonstrated a significant gap in their knowledge and experiences of data reuse. Some participants had extensive experience in reusing others' data and sharing their own data with colleagues, while others had reused data only for a few research attempts. It may lead to differences in their understanding and definition of what data reuse is. When considering data reuse, some participants may inadvertently include non-data reuse behaviors, which could ultimately affect the validity of the research conclusions. In future studies, we should adopt observational research methods, such as ethnography, to further explore IIR researchers' data reuse behaviors and to validate and expand our findings. Last but not least, the number of study participants in this study is small, but they still have a good representation of experiences, research focuses, and geographical locations.

## 7 | CONCLUSION

In this work, we aim to uncover the experiences, perspectives, and challenges faced by data reusers within the field of IIR as they attempt to reuse data in their studies. With the increasing importance of promoting data sharing and reuse within the IIR community, our work serves as an initial step towards empirically uncovering the needs and challenges faced by IR researchers, and exploring what the community need to do to overcome the challenges. We hope that this study, as well as future research, will inspire more individuals to contribute to the ongoing efforts aimed at developing a sustainable culture and policies for data reuse both within IIR and beyond, especially under the impact of AI-generated resources and generative IR. This, in turn, will foster a more seamless and effective ecosystem for sharing research data and systematically assessing data reusability, ultimately benefiting the scientific community as a whole.

## 8 | REFERENCES

# 9 | SUPPLEMENTARY INFORMATION

This document provides supplementary details about the study.

## 9.1 | Interview Guide

- Could you please provide a brief overview of your academic and research background, as well as your areas of interest and specialization in the field?

- In your research, what types of data sources do you typically rely on or collect? Are there any specific data formats or sources that you prefer working with?

- How do you go about discovering and accessing datasets that are relevant to your research within your specific domain or field? Are there any particular methods or platforms you find most helpful?

- Have you ever considered or attempted to utilize pre-existing datasets in your research projects? For instances in which you have chosen to reuse existing datasets, i.) Could you explain the advantages and tradeoffs of this approach compared to gathering new data? ii.) Could you introduce how you reuse data in your research with a case? We are especially interested in what challenges or benefits have you encountered while reusing data in your research?

- If you haven't previously explored data reuse in your research, could you provide insights into the reasons behind this decision? What obstacles or concerns have deterred you from considering data reuse as an option?

- Imagine you are browsing a data repository with the intention of finding data that could enhance your research. What criteria or information about a dataset would you consider crucial for evaluating its potential for reuse? Here are some proposed aspects to consider:

    1. Data quality and reliability
    2. Metadata completeness and accuracy
    3. Data source credibility and origin
    4. Licensing and permissions for data usage
    5. Data format and compatibility with your research tools
    6. Data recency and relevance to your research objectives
    7. Documentation and annotations accompanying the dataset

8. Imagine you are browsing a data repository with the intention of finding data that could enhanc your research, for what purpose would you try to do the data?

## 9.2 | Codebook

**TABLE 4** The Codebook

| Theme | Sub-theme | Code |
| --- | --- | --- |
| Purposes and motivations | Purposes for reusing others' data | Validate their own data or research results |
|  |  | Test applicability of their findings |
|  |  | Answer new research questions based on the data |
|  | Benefit of reusing others' data | Saving time and resources |
|  |  | Accessing data that they were not able to collect |
|  |  | Increasing the reliability of their study |
|  |  | Facilitating new collaboration opportunities |
|  | Usage of the data | Exploration |
|  |  | Ground-truthing |
| Discovering and accessing data | Discovering reusable data | Search data through public data infrastructure |
|  |  | Encounter data through academic literature |
|  |  | Learn about reusable data through personal connections |
|  | Accessing data | Obtain from personal connections |
|  |  | Access through lab website or personal repositories |
|  |  | Access though field-specific repositories |
|  |  | Ethical considerations |
| Assessing data reusability | Understandability | Logic and meanings of variables |
|  |  | Availability of documentation |
|  | Trustworthiness | Well-documented data collection procedure |
|  |  | Reputation of the data producer |
|  |  | Validity of data collection methods |
|  | Previous usage | Purpose or context of previous usage |
|  |  | Justify the data reusability |
| Concerns and challenges of reuse | Perceived data usefulness | Alignment with research goals |
|  |  | Potential for new findings |
|  | Absence of critical contextual information | Data producer failed to record during collection |
|  |  | Lack of documentation or presentation for data sharing |
|  |  | Non-standardized or inaccurate documentation |
|  | Acceptance within community | Concerns on reliability |
|  |  | Lack of innovation |